\documentclass[12pt,
]{article}

\usepackage[utf8]{inputenc}
\usepackage[T1]{fontenc}
\usepackage[english]{babel}
\usepackage[margin=3.5cm]{geometry}
\usepackage{amsmath,amssymb,amsthm}
\usepackage{mathrsfs,dsfont} %fonts
\usepackage{mathtools, ifthen, xparse, tikz-cd}
\usepackage[shortlabels]{enumitem}
\usepackage{todonotes}
\usepackage{csquotes}\MakeOuterQuote{"}
\usepackage{tikz-cd}
\usepackage[colorlinks]{hyperref}
\usepackage[capitalize]{cleveref}
\usepackage{comment}
\usepackage[misc]{ifsym}
\usepackage{multicol}

\newcommand*{\1}{\text{\usefont{U}{bbold}{m}{n}1}}
\def\RR{{\mathbb R}}
\def\CC{{\mathbb C}}
\def\NN{{\mathbb N}}
\def\B{\mathcal B}
\let\mc\mathcal

\def\K{{\mc K}}
\def\Sch{{\mathscr S}}

\def\op{\mathrm{op}^{w}} % Weyl-pseudo
\DeclareMathOperator{\id}{id}

\def\ox{\otimes}

\def\placeholder{\,\cdot\,} %{\hspace{1pt}\,{\mathbin{\vcenter{\hbox{\scalebox{0.4}{$\bullet$}}}}}\hspace{1pt}\,}} % placeholder
\renewcommand{\bar}[1]{\overline{#1}}

\DeclarePairedDelimiterX\norm[1]\lVert\rVert{\ifblank{#1}{\placeholder}{#1}}
\DeclarePairedDelimiter\abs\lvert\rvert

\DeclarePairedDelimiterX\ip[2]{\langle}{\rangle}{#1 , #2}   %SKALARPRODUKT, Hilbertraum
\DeclarePairedDelimiterX\dual[2]{\langle}{\rangle}{#1 , #2} %DUALE PAARUNG. Was kommt wo hin?

\DeclarePairedDelimiterX\braket[2]\langle\rangle{ #1 \delimsize\vert #2}
\DeclarePairedDelimiterX\ketbra[2]\vert\vert{ #1 \delimsize\rangle\delimsize\langle #2 }
\DeclarePairedDelimiterX\kettbra[1]\vert\vert{ #1 \delimsize\rangle\delimsize\langle #1 }

\newtheorem{thm}{Theorem}
\newtheorem*{thm*}{Theorem}
\newtheorem{lem}[thm]{Lemma}

\newtheorem{prop}[thm]{Proposition}

\newtheorem*{prob*}{Problem}

\theoremstyle{definition}

\theoremstyle{definition}

\theoremstyle{remark}
\newtheorem*{rem*}{Remark}
\newtheorem*{ex*}{Example}

\title{A simple criterion for essential self-adjointness of Weyl pseudodifferential operators}
\author{Robert Fulsche, Lauritz van Luijk}
\date{\today}

\begin{document}

\maketitle

\begin{abstract}
    We prove a new criterion for essential self-adjointness of pseudo\-differential operators, which does not involve ellipticity-type assumptions.
    Essential self-adjointness is proved for symbols in $C^{2d+3}$ with derivatives of order two and higher being uniformly bounded.
    These results also apply to hermitian operator-valued symbols on infinite-dimensional Hilbert spaces, which are important to applications in physics. 
    Our method relies on a phase space differential calculus for quadratic forms on $L^2(\mathbb{R}^d)$, Calder\'on-Vaillancourt type theorems and a recent self-adjointness result for Toeplitz operators.
\end{abstract}

Self-adjointness of operators is a crucial property appearing in both mathematical physics and the theory of differential equations. Weyl pseudo\-differential operators are well-known for being formally self-adjoint, provided their symbol is real-valued. Surprisingly, there are only a few criteria in the literature for extending formal self-adjointness to essential self-adjointness. Using methods recently developed in \cite{ESA}, we obtain a new criterion for essential self-adjointness of Weyl pseudodifferential operators.
The simplest version is the following:

\begin{thm}\label{thm:simple}
   Let $f\in C^{2d+3}(\RR^{2d},\RR)$ with derivatives of order $2$ to $2d+3$ being uniformly bounded, i.e., $\|\partial^\gamma f\|_\infty<\infty$ for multi-indices $\gamma\in\mathbb N_0^{2d}$ with $2\le|\gamma|\le2d+3$.
   Then $\op(f)$ is an essentially self-adjoint operator on $L^2(\RR^d)$ with domain $\Sch(\RR^d)$.
\end{thm}

We briefly comment on the assumption of bounded derivatives of second and higher orders.
In classical mechanics, the only general tool for establishing global existence of the dynamics, which is the classical analog of essential self-adjointness, is the Picard-Lindel\"of theorem \cite[Appendix to Sec.~X.I]{Reed2}.
Typically, there are two scenarios. Either the Hamiltonian function goes off to infinity in every direction so that all sub-level sets are compact and the global existence is evident, or one has to assume that the second-order derivatives are bounded.
Therefore, all symbols $f$ that satisfy the theorem also satisfy the assumptions of the Picard-Lindel\"of theorem and the converse holds up to the additional regularity imposed on the higher-order derivatives.
A naive extension of the theorem where the assumption is relaxed to, say, the third-order fails as the symbol $f(x,\xi)=\xi^2+x^3$ readily shows. Interestingly, it also fails on the classical side \cite{symptoms}. 

Note that the assumptions on $f$ imply that it is polynomially bounded, so that $f$ defines a tempered distribution on $\RR^{2d}$. This already ensures that $\op(f)$ makes sense as a continuous quadratic form on $\Sch(\RR^d)$.
Our method relies on quadratic form techniques which reduce the problem to a self-adjointness theorem for Toeplitz operators on the Segal-Bargmann space.

We indeed prove a more general result (\cref{thm:complicated}), which weakens the regularity assumptions on $f$ significantly and allows for operator-valued symbols. We also want to emphasize that this result is indeed a rather straightforward consequence of results from \cite{ESA}, in particular Theorem 20 of that paper. 
The present work should, therefore, not be seen as a paper on deep new proof ideas but rather as a report on the presented result. 
Before going into more detail, let us briefly discuss the few other results concerning essential self-adjointness of Weyl pseudodifferential operators available in the literature. 

Most papers that provide criteria for essential self-adjointness of pseuodifferential operator either assume ellipticity \cite{Nagase1988} or restrict themselves to symbols that are of a well-motivated but specific form \cite{Ichinose1993,Qihong1994}.
The paper \cite{yamazaki} due to Yamazaki also considers non-elliptic symbols and uses a more traditional approach.
Its conditions are similar in spirit but require significantly higher regularity and are obtained by completely different methods. Besides higher regularity, Yamazaki's results offer some flexibility in the choice of symbol class that the derivatives of the symbol have to be contained in. Nevertheless, besides all the flexibility, Yamazaki's theorem is not comparable to ours. For example, for symbols $b = b(x,\xi)$ (i.e., not only depending on one variable), it is always necessary that the second derivatives of $b$ satisfy some decay condition at infinity. Further, Yamazaki's theorem asks for smooth symbols, which is not necessary for our theorem. Besides all this, the flexibility in the choices of parameters does not make Yamazaki's result particularly simple in terms of applications.
Additionally, it seems that our result on essential self-adjointness is the first to deal with non-elliptic operator-valued symbols. 

We return now to our results. As mentioned above, we will allow for operator-valued symbols. 
To deal with these, we consider continuous quadratic forms on the space of vector-valued Schwartz functions.
For a general discussion of operator-valued symbols and Hilbert space-valued Schwartz spaces, we refer to \cite{Wahlberg2007}. 
If $\K$ is a Hilbert space, the vector valued Schwartz space $\Sch(\RR^d;\K)\subset L^2(\RR^d;\K)$ consists of those $L^2$-functions $\psi\in L^2(\RR^d;\K)$ such that $\ip{\psi(\placeholder)}{\xi}\in\Sch(\RR^d)$ for all $\xi\in\K$.
Alternatively, $\Sch(\RR^d;\K)$ can be identified with the tensor product $\Sch(\RR^d)\ox\K$ via $(\psi\ox\xi)(x)=\psi(x)\xi$, $x\in\RR^d)$. 
The tensor product is unique because $\Sch(\RR^d)$ is a nuclear Fr\'echet space \cite[Ch.~50]{Treves2016}. 
We will consider $\Sch(\RR^d;\K)$ with the resulting topology in the following.

% For a general discussion of operator-valued symbols and Hilbert space-valued Schwartz spaces, we refer to \cite{Wahlberg2007}. 
% We fix a separable complex Hilbert space $\K$ and denote by $\Sch(\RR^d;\K)\subset L^2(\RR^d;\K)$ the space of $\K$-valued Schwartz functions.
Given an operator-valued symbol $f \in \Sch'(\RR^{2d}; \B(\K)) := \mathcal B(\Sch(\RR^{2d}), \B(\K))$, where $\K$ is some separable Hilbert space, the Weyl pseudodifferential operator $\op(f): \Sch(\RR^d; \K) \to \Sch'(\RR^d; \K)$ defines jointly continuous quadratic form on $\Sch(\RR^d;\K)$ via
\begin{equation}\label{eq:opwf}
    (\phi,\psi)\mapsto 
    \int_{\RR^d} \int_{\RR^{d}} \int_{\RR^d} e^{i(y-x)\cdot\xi} \ip{f\big( \tfrac{x+y}{2}, \xi \big)\phi(y)}{\psi(x)} ~dy~d\xi ~dx
\end{equation}
with the natural interpretation of the above expression if $f$ is not given by an appropriately integrable function but a proper distribution \cite[Prop.~(2.5)]{Folland1989}.

Ref.~\cite{ESA} gives a sufficient condition for a hermitian operator-valued continuous quadratic form $A$ on $\Sch(\RR^d;\K)$ to define an essentially self-adjoint operator.
To state this condition, we introduce some concepts: 
Consider the usual vector $R = (Q_1,\ldots,Q_d,P_1,\allowbreak\ldots, P_d)$ of position and momentum operators on $L^2(\RR^d)$, where $Q_j\psi(x)=x_j\psi(x)$ and $P_j\psi(x)=-i\partial_j\psi(x)$. 
$R$ satisfies the canonical commutation relations $[z\cdot R,w\cdot R] = i \sigma(z,w)\1$, $z,w\in\RR^{2d}$, where $\sigma$ is the standard symplectic form on $\RR^{2d}$: 
\begin{align}
    \sigma\big((x,\xi),(y,\eta)\big) = x\cdot\eta - y\cdot\xi,\quad x,y,\xi,\eta\in\RR^d.
\end{align}
Note that we chose our conventions so that $\op(x_j)=Q_j$ and $\op(\xi_j)=P_j$.
We need the Weyl operators $W_z = e^{i \sigma(z,R)} = e^{-i(\xi\cdot Q-x\cdot P)}$, $z=(x,\xi)\in\RR^{2d}$, which define a strongly continuous projective unitary representation of $(\RR^{2d},+)$ on $L^2(\RR^d)$ which leaves $\Sch(\RR^d)$ invariant. 
The family $W_z$ indeed satisfies these properties, which is clear from the more concrete form of the Weyl operators, $W_{z}f(y) = e^{-iy\cdot\xi + \frac i2{x\cdot\xi}}f(y-x)$, where $z = (x, \xi) \in \mathbb R^{2d}$, cf.\ also \cite[p.\ 22]{Folland1989}.
In fact, $W_z\phi$ is smooth in $z$ w.r.t.\ the topology of $\Sch(\RR^d)$ for all $\phi\in\Sch(\RR^d;)$ \cite[Prop.~2]{ESA}.
In the vector-valued case, the position and momentum operators $R_j$ as well as the Weyl operators automatically make sense on $L^2(\RR^d;\K)$ as they only act on the first tensor factor in the decomposition $L^2(\RR^d;\K)\cong L^2(\RR^d)\ox\K$. 
The Weyl operators leave the vector-valued Schwartz space $\Sch(\RR^d;\K)$ invariant and smoothness of $\RR^{2d}\ni z\mapsto W_z\phi\in\Sch(\RR^d;\K)$ follows from smoothness in the scalar case.

For a continuous quadratic form $A$ on $\Sch(\RR^d;\K)$ we define the continuous form $\alpha_z(A)$ on $\Sch(\RR^d;\K)$ by $\alpha_z(A)(\phi,\psi)=A(W_{-z}\phi,W_{-z}\psi)$.
Conceptually, $\alpha_z(A)$ represents the phase space translation of $A$ by the vector $z\in\RR^{2d}$.
As an example, for the quadratic form corresponding to the operator $w\cdot R$, we get $\alpha_z(w\cdot R) = w\cdot R+(w\cdot z)\1$. 
For any pair of Schwartz functions $\phi,\psi\in\Sch(\RR^d;\K)$, $\alpha_z(A)(\phi,\psi)$ is a smooth function of $z$ and the derivatives at $z=0$ are given by 
\begin{equation}\label{eq:form_derivative}
    \partial_j \alpha_zA(\phi,\psi)\big|_{z=0} = i[\sigma(e_j,R),A](\phi,\psi) =: [\partial_j A](\phi,\psi),
\end{equation} 
where the commutator of quadratic form $A$ and a continuous linear operator $T:\Sch(\RR^d;\K)\to\Sch(\RR^d;\K)$ is the form $[T,A]=A(\placeholder,T\placeholder)-A(T\placeholder,\placeholder)$.
The so-defined quadratic forms are called the (phase space) derivatives of $A$. We define higher order derivatives $\partial^\gamma A$, $\gamma\in\NN_0^{2d}$, in a similar way (which are then given by nested commutators).

A quadratic form $A$ on $\Sch(\RR^d;\K)$ is hermitian if $A(\psi,\phi)=\overline{A(\phi,\psi)}$ for all $\psi,\phi\in\Sch(\RR^d;\K)$.
The norm $\norm A$ of a quadratic form $A$ on $\Sch(\RR^d;\K)$ is defined as the optimal constant $C>0$ such that $\abs{A(\phi,\psi)}\le C\norm\psi\norm\phi$, $\phi,\psi\in\Sch(\RR^d;\K)$ (and $\norm A=\infty$ if no such constant exists).
We now state the criterion \cite[Thm.~20]{ESA} that we want to apply.
The proof in \cite{ESA} contains a small error, which we correct in \cref{appendix}:

% \rf{in \cite{ESA}, the result was formulated using a norm which does not agree with the operator norm on $L^2(\mathbb R^d; \mathcal K)$, and the proof there does not work out when using this norm.}

% The criterion \cite[Thm.~20]{ESA} that we want to apply is formulated in terms of operator-valued quadratic form on scalar Schwartz functions but contains a small error; \rf{in \cite{ESA}, the result was formulated using a norm which does not agree with the operator norm on $L^2(\mathbb R^d; \mathcal K)$, and the proof there does not work out when using this norm.}
% The following is shown in \cref{appendix}. The idea of the proof is the same as in \cite{ESA}, but we use the correct norm and formulate the result in terms of (scalar) quadratic forms on vector-valued Schwartz functions:

\begin{lem}[{\cite{ESA}}, see \cref{appendix}]\label{thm:lemma1}
    Let $\K$ be a separable Hilbert space.
    Let $A$ be a hermitian continuous quadratic form on $\Sch(\RR^d;\K)$ satisfying
    \begin{equation}\label{eq:general_crit_forms}
        \norm{\partial_j A - \alpha_z (\partial_j A) } \le c(1 + \abs z), \quad \forall j=1,\ldots,2d,
    \end{equation}
    for some $c>0$.
    Then there is an essentially self-adjoint operator $\hat A$ on $L^2(\RR^d;\K)$ with domain $\Sch(\RR^d;\K)$ so that $\ip{\hat A\psi}{\phi}= A(\psi,\phi)$, $\psi,\phi\in \Sch(\RR^d;\K)$.
\end{lem}

In particular, the criterion is fulfilled if $\norm{\partial^\gamma A}<\infty$ for all multi-indices $\alpha$ with $\abs\alpha=2$ \cite{ESA}.
Roughly speaking, \cref{eq:general_crit_forms}\ is the condition that the first-order derivatives of $A$ have bounded oscillation.

We now return to pseudodifferential operators. As explained above, for every tempered distribution $f\in\Sch'(\RR^{2d}; \B(\K))$, the Weyl pseudodifferential operator $\op(f)$ makes sense as a quadratic form on $\Sch(\RR^d)$.
The Weyl quantization $\op$ intertwines the translations on phase space, defined by $\alpha_zf =f(\placeholder+z)$ for a function or tempered distribution $f$, with the phase space translations of quadratic forms as defined through the Weyl operators $W_z$ above:

\begin{lem}\label{lem:covariance}
    Let $f\in \mathscr S'(\RR^{2d};\B(\K))$.
    Then $\op(f)$ is a continuous quadratic form on $\Sch(\RR^d;\K)$, which satisfies
    \begin{align}\label{eq:covariance1}
        &&\alpha_z\op(f) &=\op(\alpha_z(f)), & z&\in \RR^{2d}, &&
    \intertext{and}
        &&\partial^\gamma\op(f) &=\op(\partial^\gamma f), & \gamma&\in\NN_0^{2d}. &&
        \label{eq:derivative intertwiner}
    \end{align}
\end{lem}
\begin{proof}
    \emph{Step 1:} We prove the scalar case $\K=\CC$.
    The covariance property is well known. 
    It is proved in \cite[Prop.~(2.13)]{Folland1989} for the symbol class $\Sch'(\RR^{2d})$.
    % Since we could not find a reference stating it for the class $\Sch'(\RR^{2d})$, we will provide a short proof.
    % By \cite[Sec.~3.6]{SchwartzOp}, $\op$ is an isomorphism between $\Sch'(\RR^{2d})$ and $\Sch'(L^2(\RR^d))$, where both spaces are equipped with the respective weak*-topologies.
    % The Schwartz functions  $\Sch(\RR^{2d})$ are a weak*-dense subspace of the tempered distributions $\Sch'(\RR^{2d})$.
    % That \eqref{eq:covariance1} holds for $f\in \Sch(\RR^{2d})$ is, for instance, proved in \cite[Cor.~6]{Fulsche_vanLuijk}.
    % Since the translations $\alpha_z$ act continuously on $\Sch'(\RR^{2d})$, covariance extends from $\Sch(\RR^{2d})$ to $\Sch'(\RR^{2d})$.
    % In the scalar case for $f\in\Sch$ in \eqref{eq:covariance1} is shown in \cite[Cor.~6]{Fulsche_vanLuijk} \todo{weißt du eine schöne citation oder sollen wir das beweisen? In \cite{SchwartzOp} stehts leider nicht (natürlich schon implizit).}
    % 
    % \todo{Wir k\"onnten \cite[Corollary 6]{Fulsche_vanLuijk} angeben, welches f\"ur $A \in \Sch(L^2(\mathbb R^d))$ gilt. Nach \cite{SchwartzOp} sind die Operatoren dicht in $\Sch'(L^2(\mathbb R^d))$ in w$^\ast$-Topologie, womit die Kovarianz sich auf den allgemeinen Fall \"ubertr\"agt. Ich w\"urde aber auch betonen, dass dies well-known ist und wir nur zur Vollst\"andigkeit Referenzen angeben, die thematisch zur vorliegenden Arbeit passen.}.
    We now use \eqref{eq:covariance1} to prove \eqref{eq:derivative intertwiner}. 
    For $\psi,\phi\in\Sch(\RR^d)$, let $g\in\Sch(\RR^{2d})$ be the Wigner function of $\psi\phi^*$, i.e., $g$ is the function such that $\op(g) = \psi\phi^*$, where $\psi\phi^*$ denotes the rank-1 operator $\ip{(\placeholder)}{\phi}\psi$ (in the notation of \cite[Ch.~2]{Folland1989}: $g = W(\psi,\phi)$).
    Since the Wigner function is dual to the Weyl quantization, it holds $\op(f)(\phi,\psi) = \int f(w)g(w)\,dw$, see for instance \cite[Prop.~(2.5)]{Folland1989} or \cite[Sec.~3.6]{SchwartzOp}.
    Thus, we find
    \begin{align*}
        \alpha_z\op(f)(\phi,\psi) 
        &= \op(\alpha_zf)(\phi,\psi) 
        = \int \alpha_z(f)(w)g(w)\,dw
        = \int f(w)g(w-z)\,dw.
    \end{align*}
    If we differentiate this at $z=0$ in  direction of the $j$th coordinate, the left hand-side equals $\partial_j\op(f)(\phi,\psi)$ and the right-hand side equals $-\int f(w)\partial_jg(w)\,dw = \int \partial_jf(w)g(w)\,dw = \op(\partial_jf)(\phi,\psi)$.
    Thus, the claim holds for $\abs\gamma=1$. Higher-order derivatives follow directly.

    \emph{Step 2.} The operator-valued case.
    This follows immediately from the scalar case and the fact that the involved operations act trivially on the operator part of the symbol $\op \equiv \op \ox \id_{\B(\K)}$, $\alpha_z\equiv\alpha_z\ox\id_{\B(\K)}$ and $\partial^\gamma \equiv \partial^\gamma \ox \id_{\B(\K)}$.
\end{proof}

% If we define $\alpha_z f = f(\placeholder + z)$ as the phase space shift of a function or tempered distribution, we get symbolical covariance in the sense that $\alpha_z\op(f) = \op(\alpha_z f)$, $z\in\RR^{2d}$, where the left-hand side is defined in the sense of quadratic forms as above.
% With this, one can see \cite{ESA} that $\op$ intertwines the quadratic form derivative (see \cref{eq:form_derivative}) with the distributional derivative on $\Sch'(\RR^{2d})$: 
% \todo{Der Referee hätte hier gerne eine präzise Referenz warum diese Gl.\ stimmt. Wir könnten aber auch einfach erklären wieso es so ist. Ist ja sehr zentral für das paper}
% \begin{equation}
%     \partial^\gamma \op(f) = \op(\partial^\gamma f).
% \end{equation}
% The same holds for operator-valued tempered distributions if we replace $\Sch'(\RR^{2d})$ by $\Sch'(\RR^{2d};\B(\K))= \mathcal B(\Sch'(\RR^{2d}), \B(\K))$.
With this we can restate \cref{thm:lemma1} for $A=\op(f)$ as:

\begin{lem}\label{thm:lemma2}
    Let $f\in\Sch'(\RR^{2d};\B(\K))$ be a hermitian operator-valued tempered distribution whose first-order distributional derivatives are such that 
    \begin{equation}\label{eq:general_crit_pseudo}
        \norm[\big]{\op(\partial_jf  - \partial_j f(\placeholder + z))} \le c(1+\abs z) \quad \forall j=1,\ldots,2d,
    \end{equation}
    for some $c>0$.
    Then $\op(f):\Sch(\RR^d;\K)\to L^2(\RR^d;\K)$ is essentially self-adjoint.
    In particular, this holds if the second-order derivatives satisfy $\norm{\op(\partial_i\partial_j f)}<\infty$ for all $i,j=1,\ldots,2d$.
\end{lem}

The condition in Eq.~\eqref{eq:general_crit_pseudo}, which is essentially a bound on the oscillation of $\nabla f$, can now be checked using Calder\'on-Vaillancourt type theorems.
With the theorem \cite[Thm.~2.73]{Folland1989} we obtain \cref{thm:simple} as a corollary of \cref{thm:lemma2} and with its operator-valued version \cite[Thm.~A.6]{teufel} we obtain the analog of \cref{thm:simple} for operator-valued symbols. One can use more general symbol classes that yield bounded Weyl pseudodifferential operators, such as Sj\"{o}strand's class $M_{\infty, 1}$. We do not want to give the precise definition of this class here; instead, we refer to the literature, e.g.\ \cite[Sec.~3]{Sjostrand1994} or \cite[Thm.~1.1]{Grochenig_Heil1999} for the scalar-valued and \cite[Cor.~4.9]{Wahlberg2007} for the operator-valued case. Instead, we simply want to emphasize that membership in $M_{\infty, 1}$ does not need any form of differentiability (even though functions in $M_{\infty, 1}$ are always continuous). Further, $M_{\infty, 1}$ contains some well-known symbol spaces such as the Calder\'{o}n-Vaillancourt class and the H\"{o}lder-Zygmund classes $\Lambda^s(\RR^{2d})$ for $s > 2d$, see \cite[Prop.~3.6]{Grochenig_Heil1999}. 
Applying the Calder\'on-Vaillancourt theorem for $M_{\infty, 1}$ symbols yields our final result:

\begin{thm}\label{thm:complicated}
    Let $\K$ be a separable Hilbert space.
    Let $f\in  \Sch'(\RR^{2d};\allowbreak\B(\K))$ be a hermitian operator-valued symbol such that the $\partial^\alpha f \in M_{\infty, 1}(\RR^{2d};\allowbreak \B(\K))$ for all $|\alpha| = 2$. 
    Then $\op(f)$ is an essentially self-adjoint operator on $L^2(\RR^d; \K)$ with domain $\Sch(\RR^d; \K)$.
\end{thm}

\begin{rem*}
    The interested reader might wonder if our main result is possibly also obtainable as a consequence of Nelson's commutator theorem \cite[Thm.\ X.37]{Reed2}, by comparing with the harmonic oscillator. Indeed, this is one ingredient used in the proof of Lemma \ref{thm:lemma1} in \cite{ESA}. Given the fact that the Weyl quantization is not order-preserving, it seems necessary to the authors to pass from $\op(f)$ to $\op(\widetilde{f})$ when wanting to apply the commutator theorem, where $\widetilde{f}$ is the \emph{heat transform} of $f$ at a suitable time. The analysis of this is exactly what has been done in \cite{ESA}. 
    Nevertheless, there the main result was formulated as a result on Toeplitz operators on the Segal-Bargmann space. Since such operators are unitarily equivalent to Weyl pseudodifferential operators with heat-transformed symbols, the results of \cite{ESA} can also be understood as results on Weyl pseudodifferential operators. The phase space formalism presented there and also in this paper should not be seen as a necessity to obtain our main result, but as a very convenient framework to work within.
\end{rem*}

\subsection*{Acknowledgements.}
LvL acknowledges funding by the MWK Lower Saxony (Stay Inspired Grant 15-76251-2-Stay-9/22-16583/2022). Both authors acknowledge the reviewer's valuable comments on the present work.

{\small \noindent
\textbf{Author Contributions} Both authors contributed equally and reviewed the manuscript.

\noindent \textbf{Data availability} No datasets were generated or analysed during the current study.

\noindent \textbf{Conflict of interest} The authors declare no competing interests.
}

\newpage

\appendix
\section{Corrected proof of Theorem 20 in \texorpdfstring{\cite{ESA}}{[1]}}\label{appendix}

In \cite[Thm.~20]{ESA}, the self-adjointness criterion is formulated for continuous operator-valued quadratic forms on the space of Schwartz functions.
As in the main text, we use (scalar-valued) quadratic forms on vector-valued Schwartz function instead.
Although both notions are, in principle, equivalent, the latter is much easier to work with.

We briefly explain what went wrong in \cite{ESA}:
% It seems that the main reason for the error is the usage of operator-valued forms.
The norm that appears in \cite[Thm.~20]{ESA} and its proof should correspond to the operator norm for operators on $L^2(\RR^d;\K)$ but is defined falsely, likely due to the usage of operator-valued forms.
The idea of the proof is to show essential self-adjointness by showing that the given form is a bounded perturbation of a certain Toeplitz operator, which is essentially self-adjointness by the main theorem of \cite{ESA}.
Although the argument in the proof itself is correct, it fails to show essential self-adjointness because the finiteness of the falsely defined norm does not imply boundedness.

We show the following:
% \rf{in \cite{ESA}, the result was formulated using a norm which does not agree with the operator norm on $L^2(\mathbb R^d; \mathcal K)$, and the proof there does not work out when using this norm.}

% The criterion \cite[Thm.~20]{ESA} that we want to apply is formulated in terms of operator-valued quadratic form on scalar Schwartz functions but contains a small error; \rf{in \cite{ESA}, the result was formulated using a norm which does not agree with the operator norm on $L^2(\mathbb R^d; \mathcal K)$, and the proof there does not work out when using this norm.}
% The following is shown in \cref{appendix}. The idea of the proof is the same as in \cite{ESA}, but we use the correct norm and formulate the result in terms of (scalar) quadratic forms on vector-valued Schwartz functions:

\begin{prop}[{\cite[Thm.~20]{ESA}}]\label{prop:correction}
    Let $A$ be a hermitian continuous quadratic form on $\Sch(\RR^n;\K)$.
    Suppose that the derivatives of $A$ have bounded oscillation in the sense that there is a $c>0$ such that
    \begin{equation}
        \|\partial_jA - \alpha_z(\partial_jA)\| \le c (1+|z|) \qquad \forall j=1,\ldots,2d,\ z\in\RR^{2d}.
    \end{equation}
    Then there is an essentially self-adjoint operator $\hat A$ with domain $\Sch(\RR^d;\K)$ so that $\ip{\hat A\psi}{\phi}_{L^2(\RR^d;\K)} = A(\psi,\phi)$ for all $\psi,\phi\in\Sch(\RR^d;\K)$.
\end{prop}

% We work in the Segal-Bargman(n) representation, where $L^2(\RR^d;\K)$ is identified with the closed subspace of complex analytic functions of $L^2(\CC^d,\mu;\K)$, where $\mu$ is the Gaussian measure $d\mu(a) = e^{-|a|^2/?}da$.\todo{die punkte im reellen phasenraum heißen $z$ darum nenne ich die im complexen jetzt $a$. können aber auch die notation ändern.}
% In order to describe the isomorphism explicitly, we introduce the family of \emph{coherent states} $\psi_z\in L^2(\RR^d)$, $z\in\RR^{2d}$.
% For $z=0$, we have $\psi_0(x) = ??? e^{-|x|^2/ ?}$ and for general $z$, we have $\psi_z=W_{\pm z}\psi_0$.
% Then, the isomorphism between $L^2(\RR^d;\K)$ and the $\K$-valued Segal-Bargman(n) space is given by
% \begin{equation}
%     U\psi (a) \approx \ip{\psi}{k_{z(a)}}  ? ???,
% \end{equation}
% where $z(a) = ???$.

We need to work with the family of \emph{coherent states} $\psi_z\in L^2(\RR^d)$, $z\in\RR^{2d}$.
For $z=0$, we have $\psi_0(x) = \pi^{-d/4} e^{-|x|^2/2}$ and for general $z$, we have $\psi_z=W_{z}\psi_0$, where $W_z$ denotes the Weyl operators, introduced in the main text.
For a continuous quadratic form $A$ on $\Sch(\RR^d;\K)$ and $z\in\RR^{2d}$, we have a norm-continuous quadratic form $(\xi,\eta)\mapsto A(\psi_z\ox\xi,\psi_z\ox\eta)$ on $\K$.
We denote the bounded operator implementing this form by $\tilde A(z)\in\B(\K)$ and refer to it as the \emph{Berezin transform} of $A$ at $z$.
Note that the Berezin transform defines a continuous function $\tilde A : \RR^{2d}\to \B(\K)$.

For a polynomially bounded measurable symbol $f : \RR^{2d}\to \B(\K)$, the Toeplitz operator $T_f$ is most naturally defined as an operator on the Segal-Bargmann-Fock space. For our purposes, the most useful representation of $T_f$ is given by
\begin{equation}\label{def:Toeplitz}
    T_f = \frac{1}{(2\pi)^d} \int_{\mathbb R^{2d}} P_z \ox f(z)~dz,
\end{equation}
where $P_z$ denotes the orthogonal projection onto the subspace spanned by $\psi_z$.
The integral in \eqref{def:Toeplitz} should be understood weakly, i.e., as a quadratic form on $\Sch(\RR^d; \K)$. This representation of a Toeplitz operator is unitarily equivalent to the standard form on the Segal-Bargmann-Fock space by means of the Bargmann transform. We refer to \cite{Folland1989, Zhu2012} regarding details on this transform.

For the proof, we need the following Lemmas:

\begin{lem}[{\cite[Lem.~16]{ESA}}]
    Let $F:\RR^n\to \CC$ be a $C^1$-function whose gradient $\nabla F$ has bounded oscillation in the sense that $|\partial_j F(x) -\partial_j F(y)| \le c(1+|x-y|)$ for some $c>0$ and all $x,y\in \RR^n$.
    Then the function $R_x:\RR^n\to\CC$ defined for $x\in\RR^n$ by
    \begin{equation}
        R_x(y) = F(y+x)- F(y)-x\cdot \nabla F(y)
    \end{equation}
    is uniformly bounded with $|R_x(y)| \le c(|x|+|x^2|)$ for all $y\in\RR^n$.
\end{lem}

We need injectivity of the Berezin transform, which is well-known in the scalar-valued case. We only sketch the well-known proof.

\begin{lem}\label{lem:Berezin injective}
    The Berezin transform $A\mapsto \tilde A$ is injective on the space of continuous quadratic forms on $\Sch(\RR^d;\K)$.
\end{lem}
\begin{proof}
    The standard proof of injectivity for the case of bounded operators on scalar-valued $L^2$-functions, or equivalently on the Segal-Bargmann space \cite{Zhu2012}, generalizes to our setting:
    By linearity, it suffices to show that $\tilde B =0$ implies $B=0$ for a continuous quadratic form $B$ on $\Sch(\RR^d;\K)$.
    Indeed, if $\tilde B = 0$ is equivalent to $B(\psi_z\ox\xi,\psi_z\ox\eta)=0$ for all $z\in\RR^{2d}$.
    If we identify $\RR^{2d}=\CC^d$ such that the symplectic form $\sigma$ on $\RR^{2d}$ is identified with the imaginary part of the inner product on $\CC^d$, then $z\mapsto \psi_z$ is a complex analytic function.
    Thus, $(z,w)\mapsto B(\psi_{\bar z}\ox\xi,\psi_w\eta)$ is complex analytic for all $\xi,\eta\in\K$.
    By the identity theorem in \cite[Proposition 1.69]{Folland1989} (or rather, its straightforward vector-valued analogue), $\tilde B=0$ implies that $B$ vanishes on all coherent states, i.e., $B(\psi_z\ox\xi,\psi_w\ox\eta)=0$ for all $\xi,\eta\in\K$. Since such elements span a dense subspace of $\Sch(\RR^{d}; \mathcal K)$, it follows that $B(f, g) = 0$ for every $f, g \in \Sch(\RR^d; \mathcal K)$.
\end{proof}

\begin{lem}\label{lem:TtildeA}
    Let $A$ be a continuous quadratic form on $\Sch(\RR^d;\K)$.
    Then the Topelitz operator whose symbol is the Berezin transform $\tilde A$ of $A$ is given by the weak integral
    \begin{equation}\label{eq:TtildeA}
        T_{\tilde A} = (2\pi)^{-d}\int_{\RR^{2d}} \alpha_w(A)e^{-\frac{|w|^2}2}~dw.
    \end{equation}
\end{lem}
\begin{proof}
    It is easy to check that the Berezin transforms of both sides of \eqref{eq:TtildeA} coincide.
    The claim then follows from \cref{lem:Berezin injective}.
\end{proof}

\begin{proof}[Proof of \cref{prop:correction}]
    % The idea of the proof is the same as in \cite{ESA}: We show that $A$ differs from $T_{\tilde A}$ by a bounded quadratic form and use that the Toeplitz operator $T_{\tilde A}$ is essentially self-adjoint by the main theorem of \cite{ESA} to conclude that $A$ is essentially self-adjoint.
    We define a continuous quadratic form $R_w$ on $\Sch(\RR^d;\K)$ by
    \begin{equation*}
        R_w = \alpha_w A - A - w\cdot \nabla A,
    \end{equation*}
    where $w\cdot\nabla A=\sum_j w_j \partial_jA$.
    We begin by showing
    \begin{equation}\label{Eq:2}
        A- T_{\tilde A} = (2\pi)^{-d} \int R_w\, e^{-\frac{|w|^2}{2}}dw,
    \end{equation}
    where the integral is taken in the weak sense. 
    % For this identity, we first observe that we can write
    % \begin{align*}
    %     T_{\widetilde{A}} = (2\pi)^{-d}\int_{\RR^{2d}} \alpha_w(A) e^{-\frac{|w|^2}{2}}~dw,
    % \end{align*}
    % which can be verified by comparing the Berezin transforms of both sides (the Berezin transform is injective). 
    Indeed, Eq.~\eqref{Eq:2} follows from \cref{lem:TtildeA} and $\int w_j e^{-|w|^2/2}dw =0$:
    \begin{align*}
        A-T_{\tilde A} &= (2\pi)^{-d} \int_{\mathbb R^{2d}} (A- \alpha_wA)  e^{-\frac{|w|^2}{2}}dw \\
        &= (2\pi)^{-d} \int_{\RR^{2d}} R_w e^{-\frac{|w|^2}{2}}dw + \sum_{j=1}^{2d}\int_{\RR^{2d}} w_j e^{-\frac{|w|^2}{2}}dw\, \partial_jA\\
        &= (2\pi)^{-d} \int_{\RR^{2d}} R_w e^{-\frac{|w|^2}{2}}dw.
    \end{align*}
    
    We show that $(2\pi)^{-d}\int_{\RR^{2d}} R_w\, e^{-\frac{|w|^2}{2}}dw$ is a bounded quadratic form.
    For $f,g\in\Sch(\RR^d;\K)$, the function $F(w) = \alpha_wA(f,g)$ is differentiable and $|\partial_j F(w)-\partial_j F(z)| \le \norm f\norm g c(1+|z-w|)$ with $c>0$ due to the assumptions of the proposition.
    Thus, the lemma above implies
    \begin{equation*}
        |R_w(f,g)| = |F(w)-F(0)-w\cdot \nabla F(0)| \le c\norm f\norm g (1+\abs w).
    \end{equation*}
    Therefore, $A-T_{\tilde A} = (2\pi)^{-d}\int_{\RR^{2d}} R_w\, e^{-\frac{|w|^2}{2}}dw$ is a bounded quadratic form.
    Since, by the main theorem of \cite{ESA}, $T_{\tilde A}$ is an essentially self-adjoint operator on $\Sch(\RR^d;\K)$, the claim follows.
\end{proof}

\bibliographystyle{abbrv}
\bibliography{refs}

\vspace{1cm}

\begin{multicols}{2}

\noindent
Robert Fulsche\\
\href{fulsche@math.uni-hannover.de}{\Letter ~fulsche@math.uni-hannover.de}
\\
\noindent
Institut f\"{u}r Analysis\\
Leibniz Universit\"at Hannover\\
Welfengarten 1\\
30167 Hannover\\
GERMANY\\

\noindent
Lauritz van Luijk\\
\href{lauritz.vanluijk@itp.uni-hannover.de}{\Letter ~lauritz.vanluijk@itp.uni-hannover.de}
\\
\noindent
Institut f\"{u}r Theoretische Physik\\
Leibniz Universit\"at Hannover\\
Appelstra\ss e 2\\
30167 Hannover\\
GERMANY

\end{multicols}
\end{document}